# ProTCT: Projection quantification and fidelity constraint integrated deep reconstruction for Tangential CT

Bingan Yuan, Bowei Liu, Kang Fan, Zheng Fang*

*School of Aerospace Engineering, Xiamen University, Xiamen 361102, China*

## ABSTRACT

Computed tomography (CT) is a useful tool for non-destructive testing of large-diameter samples, such as oil pipelines and rockets. However, the projections are truncated along the detector direction, resulting in degraded images with radial artifacts. Meanwhile, existing methods fail to achieve satisfactory reconstruction quality because of the ill-defined sampling conditions in the projection domain and over-smoothing in the image domain. In this paper, we propose a Projection quantification and fidelity constraint integrated deep reconstruction method for Tangential CT (ProTCT) to improve the reconstruction quality for large-diameter object imaging. Specifically, the sampling conditions for reconstruction are analyzed, offering practical guidelines for TCT firmware design. In addition, a deep artifact-suppression network and a fidelity-constraint module are introduced to operate across both projection and image domains to remove artifacts and restore edge details. As demonstrated on simulated and real datasets, ProTCT demonstrates good performance in structure restoration and detail retention. This work advances the understanding of the sampling conditions and the reconstruction methods of TCT, further promoting the application of CT in industrial inspection.

*Keywords*: Tangential computed tomography (TCT), Projection quantification, Fidelity constraint, Cross-domain reconstruction.

## 1. Introduction

Tangential CT (TCT) is a special CT imaging modality in which the projections are taken along lines outside the central areas. It was first introduced for the medical diagnosis of the external chest cavity, avoiding the beating heart [1], and then expanded to industrial non-destructive testing for large tubular samples, such as rockets and oil pipelines [2,3]. However, TCT reconstruction is a severely ill-posed problem because the measured projections are truncated in the detector direction. The classical filtered back-projection (FBP) reconstruction algorithm, which requires full-beam coverage, produces degraded images in this situation. Over the past decades, numerous approaches to TCT reconstruction have been developed. These methods can be broadly divided into projection-domain sampling strategies and image-domain regularized reconstruction methods. In the projection domain, existing methods mainly aim to ensure sufficient data sampling by adjusting the projection coverage or compensating for truncated measurements. Schafer et al. [4] stated that scanning half of the transverse extent of the object provides sufficient sampling in a circular trajectory. However, the exact positioning of the half-scan is difficult in practical scenarios, so the obtained projections are usually beyond half of the object, and the redundant data need to be addressed [5]. To further reduce the detector coverage, Quinto et al. [6,7] investigated sub-half-scan configurations by extrapolating the truncated projections to the half-scan condition. Overall, these projection-domain approaches improve TCT reconstruction by adjusting the measured or extrapolated projection coverage to meet predefined sampling requirements. However, they generally rely on empirically or geometrically specified coverage conditions, while the necessary projection coverage for different objects and scanning configurations remains unclear.

* Corresponding author. School of Aerospace Engineering, Xiamen University, Xiamen 361102, China.
*E-mail address*: fangzheng@xmu.edu.cn

In the image domain, regularization-based iterative reconstruction methods have been developed to exploit the directional characteristics of TCT artifacts. Guo et al. [8] found that the artifacts in TCT were mainly distributed along the radial direction, and a weighted directional total variation (WDTV) algorithm was proposed to enhance the radial edges. Zeng et al. [9] divided the image into several subregions using the Chan-Vese level set method, and then applied a projection onto convex sets (POCS) with a TV minimization reconstruction algorithm. Qin et al. [10] proposed an edge-preserving diffusion and smoothing regularization model, which could suppress both truncation and beam hardening artifacts. Shen et al. [11] developed an anisotropic relative total variation in polar coordinates (P-ARTV) model to suppress the artifacts. Although these handcrafted regularization methods effectively exploit the directional characteristics of TCT artifacts, their strong smoothing priors may suppress fine structures and lead to the loss of image details.

With the rapid development of deep learning (DL), data-driven reconstruction methods have been extensively investigated for incomplete and degraded CT imaging, including limited-angle, sparse-view and low-dose CT. These methods can generally be categorized into projection-image dual-domain networks, model-driven reconstruction networks, and learned-prior-based reconstruction networks. For dual-domain reconstruction, Hu et al. [12] proposed two U-Net architectures to complete missing projections and suppress image artifacts. Wu et al. [13] developed a dual-domain network for sparse-view CT, which refined image details based on residuals between images and projections. Such methods exploit complementary information from the projection and image domains and have demonstrated advantages over purely image-domain post-processing. For model-driven or unrolled reconstruction, which incorporates CT imaging operators into neural networks to improve data consistency and interpretability. Zhang et al. [14] introduced an unrolled deep residual network to estimate the residual image from projection errors. Long et al. [15] presented a prunable UNET++ network with hybrid attention and a composite loss function, which jointly emphasized noise characteristics and the sparsity of image gradients. Xia et al. [16] proposed a learnable local-nonlocal regularization model, where a neural network was employed to replace the gradient of handcrafted regularization and transformer-based encoders and decoders were introduced to incorporate nonlocal priors into the reconstruction framework. More recently, advanced generative and adaptive learning strategies have also been explored. Gao et al. [17] developed a contextual error-modulated generalized diffusion model for low-dose CT denoising, incorporating contextual priors into the sampling process and improving generalizability through a one-shot learning strategy. Yang et al. [18,19] proposed hypernetwork-based physics-driven personalized federated learning methods to enable institution- or scanner-specific adaptation in low-dose CT imaging. Despite these advances, existing DL-based methods have mainly focused on predefined degradation patterns, such as limited-angle, sparse-view, or low-dose CT. The projection truncation in TCT is different because the necessary projection coverage is closely related to the object geometry and scanning configuration, leading to object-dependent sampling conditions.

Inspired by all the work above, this paper proposes a Projection quantification and fidelity constraint integrated deep reconstruction framework for TCT (ProTCT). The core novelty of this work lies in the joint optimization of the TCT scanning scheme and reconstruction strategy. The main contributions are summarized as follows:

1) A projection scanning scheme grounded in TCT imaging physics is developed to characterize the necessary sampling conditions. By jointly considering angular coverage, projection views, and reconstruction stability, the proposed sampling model determines the necessary projection conditions and detector extension length.

2) A projection fidelity constraint is incorporated into the cross-domain reconstruction strategy to introduce physically measured projection information. The residual between the reprojected network output and the measured projections is used to enforce measurement consistency and preserve structural details.

3) A coarse-to-fine deep reconstruction architecture is designed. The cascaded artifact-suppression and image-refinement networks employ a Swin-Transformer-enhanced encoder-decoder and asymmetric convolution blocks to capture nonlocal features and suppress characteristic radial and secondary artifacts in TCT.

The rest of this paper is organized as follows. Section 2 introduces the details of the ProTCT framework. Section 3 presents the simulated and real data results, and analyzes the mechanisms underlying the proposed method. Section 4 concludes the paper and discusses several related issues.

## 2. Methodology

First, a summary of the method is presented. Without loss of generality, we consider TCT scanning with annular object, as shown in Fig. 1. Limited by the size of the detector, only a portion of the object can be measured in a single view. In the extreme case, inner-circle points (e.g., point *A* in Fig. 1) can be projected only once during the circular scanning trajectory. If a conventional analytical reconstruction method is applied, the reconstructed image will contain radial artifacts and the inner regions will be heavily distorted, as shown in Fig. 2.

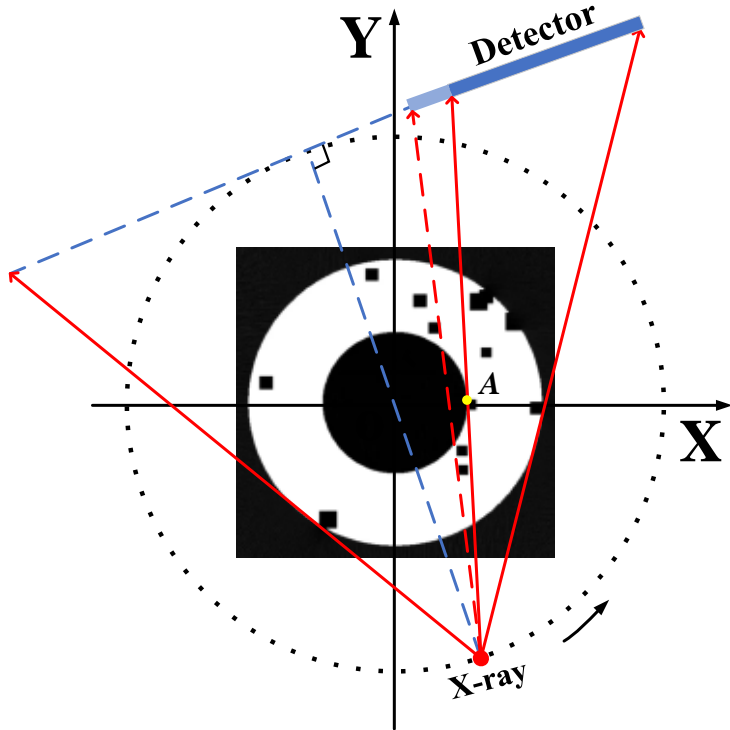

Fig. 1. Illustration of tangential CT scanning.

Hence, this manuscript explores the necessary (or minimal) projections required for reconstructing each point in CT, namely, by calculating the required detector extension length (e.g., the red dotted line in Fig. 1 for point A), such that the necessary projections within a limited angular range (as shown in Fig. 5) are physically measured by the detector. This analysis provides guidance for the practical firmware design of the detector. However, due to the intrinsic detector truncation in TCT, characteristic radial artifacts still appear in the reconstructed images when conventional analytical or iterative reconstruction methods are applied. Therefore, this observation motivates a reconstruction framework that combines physical-consistent projection information with data-driven compensation.

In this paper, we propose a specialized reconstruction framework (ProTCT), as shown in Fig. 3. The ProTCT framework consists of four cascaded stages. The first stage (Fig. 3(a)) involves quantifying the necessary projections for TCT reconstruction and obtaining an initial image. The second stage (Fig. 3(b)) employs an artifact suppression network (IAS-Net) to reduce the artifacts in the initial image. The third stage (Fig. 3(c)) incorporates the quantified projections to constrain the output of IAS-Net via a projection fidelity constraint module (PFC-Module). The final stage (Fig. 3(d)) involves the image refinement network (IR-Net) to refine the result of the PFC-Module.

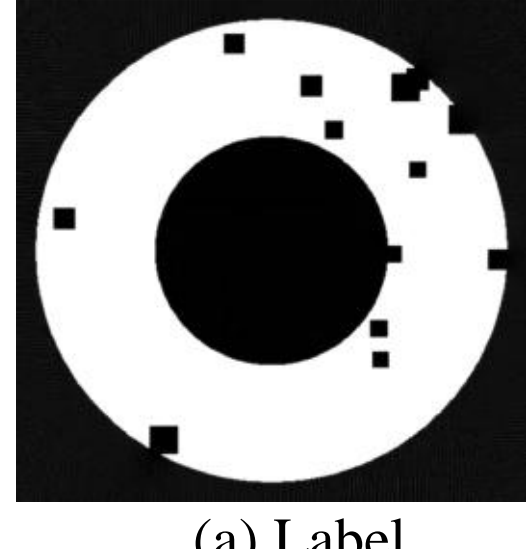

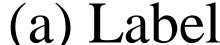

(a) Label

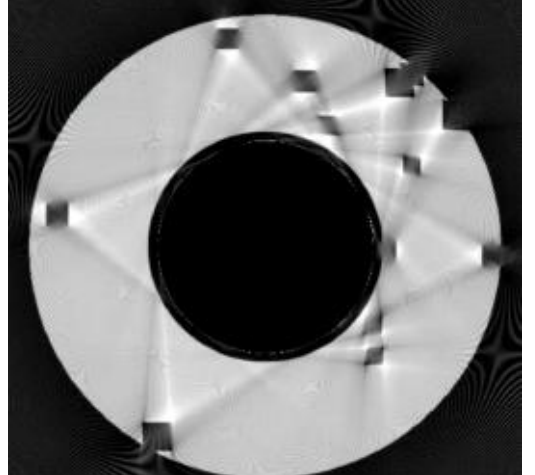

(b) FBP reconstruction of TCT

Fig. 2. Illustration of the distorted reconstruction of TCT.

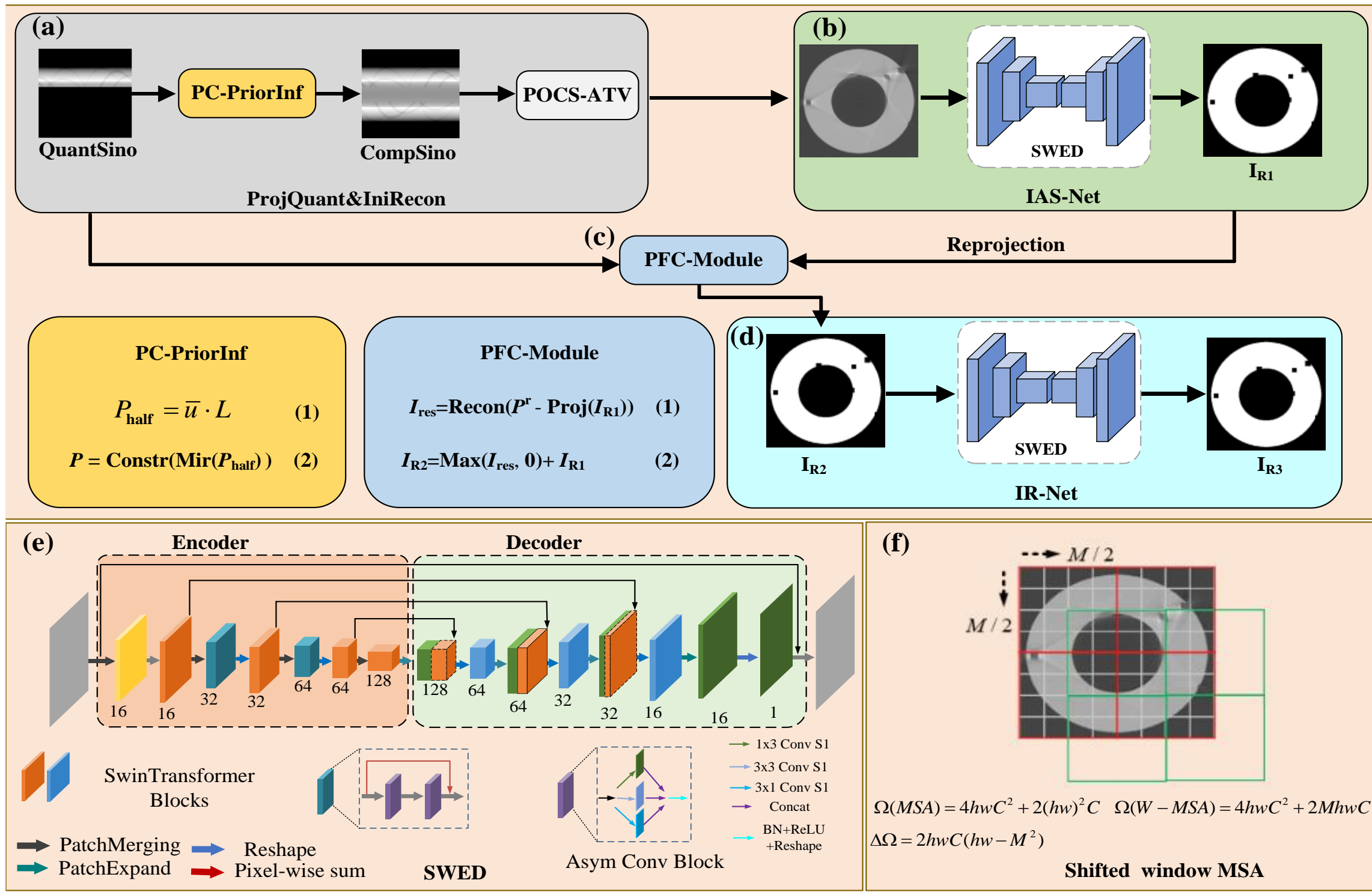

Fig. 3. The framework of ProTCT. (a) The projection quantification and initial reconstruction of TCT; (b) The structure of Image Artifact Suppression Network (IAS-Net); (c) The process of Projection Fidelity Constraint module (PFC); (d) The structure of Image Refinement Network (IR-Net); (e) The structure of SWin-transformer-enhanced Encoder-Decoder (SWED), which is the backbone of IAS-Net and IR-Net. (f) The shifted local window strategy of SWED.

### *2.1. Projection quantification with ATV Regularization*

The first part of ProTCT (Fig. 3(a)) focuses on projection quantification and the initial reconstruction of TCT. During the scanning process of TCT, each point receives a different set of projections, characterized by its angular coverage ($\theta$) and number of views ($N_{view}$), as illustrated in Fig. 4.

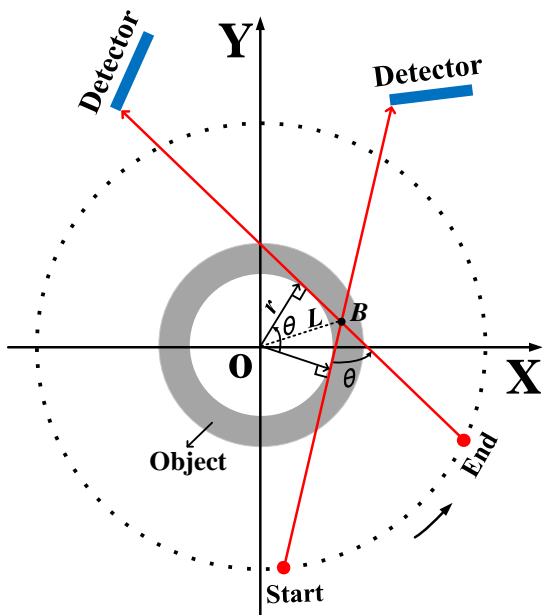

Fig. 4. The obtained scanning angle of a point in TCT.

As shown in Fig. 4, the scanning angle $\theta$ increases with the radius $L$ of the measured point, as follows:

$$\theta = 2\arccos(\frac{r}{L}) \tag{1}$$

where $r$ is the inner-circle radius of the annulus, $L$ is the radius of the measured point.

It is evident that inner-circle points receive fewer projections, making them suitable candidates for investigating the necessary projection conditions in TCT reconstruction. Taking point $C$ in Fig. 5 as an example, to increase its angular coverage $\theta$, the detector can be extended by a length $d'$, which results in a tangential ray tilting by an additional distance of $d$. Under this setup, point $C$ is uniformly sampled in a limited angular range of $\theta$, corresponding to $N_{view}$ views.

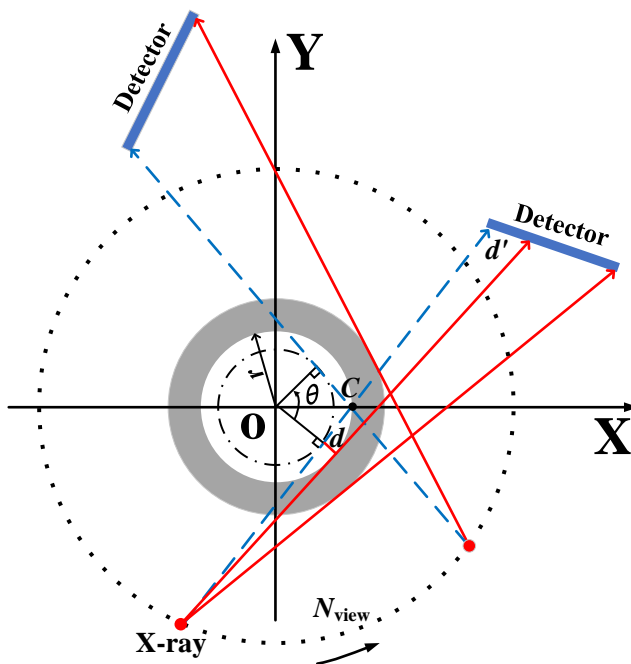

Fig. 5. The inner-circle point scanning process in TCT.

The relationships between $\theta$, $d$, and $d'$ can be expressed by [20]:

$$\theta = 2\arccos\frac{r-d}{r} \tag{2}$$

$$d' = \frac{r \cdot \mathrm{SDD}}{\sqrt{\mathrm{SOD}^2 - r^2}} - \frac{(r-d)\cdot \mathrm{SDD}}{\sqrt{\mathrm{SOD}^2 - (r-d)^2}} \tag{3}$$

where SDD and SOD denote the source-to-detector distance and source-to-rotation-center distance, respectively.

From Eqs. (2)-(3), once the scanning angle $\theta$ is determined, the corresponding detector extension length can be derived. This relationship provides a theoretical foundation for the firmware design of the TCT system.

To quantify the necessary projection $(\theta, N_{\mathrm{view}})$ for image reconstruction of TCT, we refer to the sufficient and necessary sampling conditions for limited-angle CT (LACT) described in [21]. The scanning process of LACT is illustrated in Fig. 6. Under the assumption of an individual ray (line-integral) imaging model and neglecting scattering effects, the projection views acquired for each point in LACT are equivalent to those of the corresponding point in TCT within a limited angular range, which allows us to adopt the necessary projection conditions for projection quantification. For a specific inner-circle point (e.g., point $D$ in Fig. 6), the acquired projection (i.e., uniform sampling $N_{view}$ over a limited angular range $\theta$) is equivalent to that in TCT for the corresponding point (e.g., point $C$ in Fig. 5). In other words, the necessary projection requirements for reconstructing each point in LACT are also applicable to TCT, based on which the detector extension length can be calculated using Eqs. (2)-(3). However, the sufficient projection conditions in LACT are not applicable to TCT due to the differences in imaging geometry. The reconstruction of LACT is based on the anisotropic total variation (ATV) regularization model, as:

$$\min_x R(x), \; s.\,t.\; Ax = b \tag{4}$$

$$R(x) = \| \mathrm{D}^T x \|_1 = \sum_{i=1}^{N} | d_i^T x | \tag{5}$$

where $A \in R^{m\times n}$ is the sampling system matrix, and $m = N_{view} \cdot N_{bin}$ is the number of measurements, $N_{bin}$ is the number of detector units, $n$ is the number of image pixels. The term $d_i^T x$ represents the horizontal or vertical gradients of a pixel, $N = 2n$, and $D \in R^{n\times N}$.

The properties of the system matrix $A$ are determined by the projection $(\theta, N_{\mathrm{view}})$, and the necessary and sufficient conditions on $A$ to solve the ATV model are given in [22]:

1) The matrix $(A; D_{I^C}^T)$ has full rank.
2) The minimum value of the variable $t$ under the following constraints is less than 1:

$$\min_{t,v,w}(t)\text{, s. t. } \begin{cases} -t\mathbf{1} \le v_{I^C} \le t\mathbf{1} \\ A^T w = D_I v_I + D_{I^c} v_{I^c} \\ v_I = sign(D_I^T x) \end{cases} \tag{6}$$

where $w \in R^m$, $v \in R^N$, $I = \text{supp}(x)$, and $I^C$ is the complement of the index set $I$, $D_I$ is the submatrix of $D$ with columns indexed by $I$, the vector $v_I$ is the sub-vector of $v$ with rows indexed by $I$.

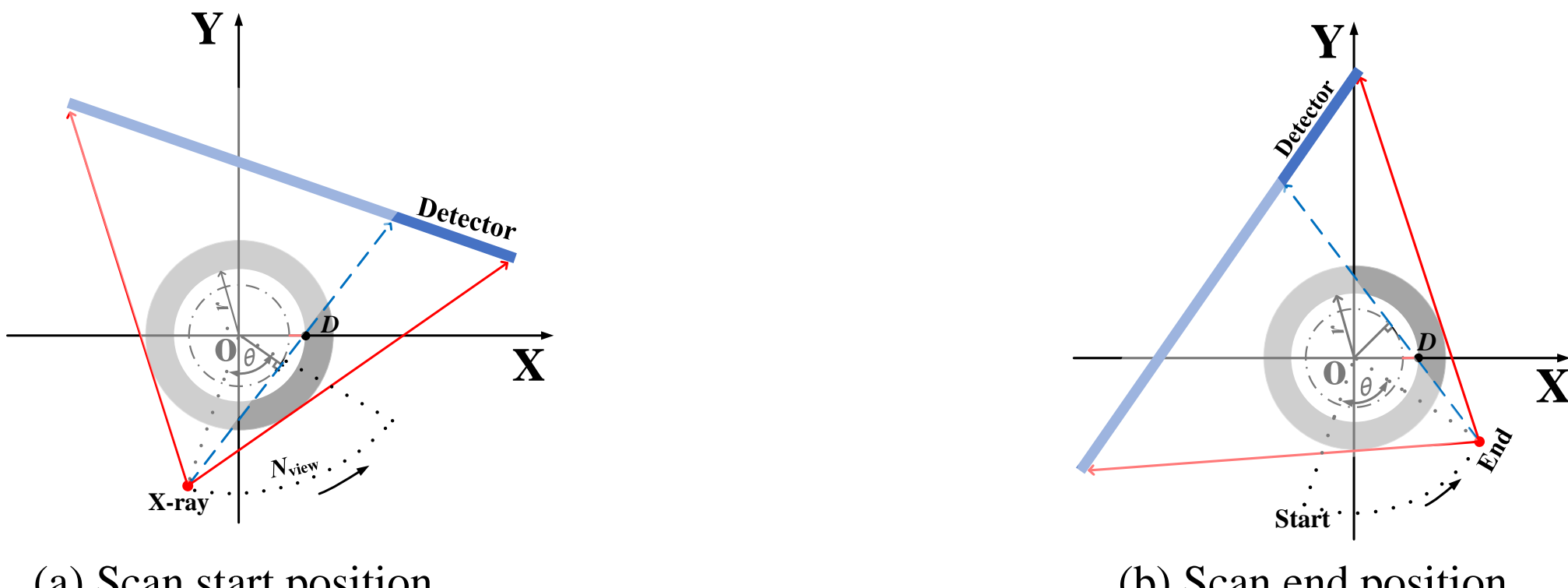

(a) Scan start position (b) Scan end position

Fig. 6. The scanning process of the inner circle point in limited-angle CT.

It is worth noting that the variable $t$ in the second condition (Eq. (6)) has no direct physical meaning, its detailed derivation can be found in [22]. The first condition, the matrix $(A; D_{I^c}^T)$ has full rank, is easily satisfied for the ATV minimization model. For the second condition, one can first set an angle range $\theta$, and then search for suitable views $N_{\text{view}}$ corresponding to different projection sets $(\theta, N_{\text{view}})$ (which alter the properties of matrix $A$). If the projection sets to the linear optimization problem in Eq. (6) satisfies $t$<1, the ATV model has a unique solution; that is, the image can be reconstructed exactly.

However, different solutions $(\theta, N_{\text{view}}, t)$ to the optimization problem will have varying implications for the TCT system design. A smaller $\theta$ can reduce the required expansion of the detector (as illustrated in Fig. 5), but this results in more sampling ($N_{view}$), meaning more data must be acquired and the servo system must operate with higher precision. Moreover, the larger $\theta$ and $N_{view}$ are, the smaller $t$ becomes, indicating improved reconstruction stability. In essence, the parameters in $(\theta, N_{\text{view}}, t)$ are mutually constrained and exhibit trade-offs. To address these interdependencies, we propose a sampling model for TCT as follows:

$$\min\left(\left(\frac{\theta}{T_\theta}\right)^a \cdot \left(\frac{N_{\text{view}}}{T_{N_{\text{view}}}}\right)^b \cdot \left(\frac{t}{T_t}\right)^c\right) \tag{7}$$

where $T_\theta$, $T_{N_{view}}$ and $T_t$ denote the maximum values of angle range $\theta$, number of views $N_{\text{view}}$ and variable $t$, respectively. These thresholds can be selected according to practical imaging requirements.

The sampling model (Eq. (7)) of TCT is constructed based on the following principles: (1) each variable is normalized to ensure that differences in units or scales do not dominate the model, allowing for a balanced contribution from all variables; (2) individual weights are assigned to each variable, enabling flexible adjustment of their penalty levels or importance in the decision process; (3) a multiplicative structure is used to correlate the variables, effectively capturing the nonlinear interactions among them.

*2.2. Projection completion and initial reconstruction*

The necessary projection data for TCT can be quantified using the sampling model introduced in Section 2.1. However, because the projection acquired in each view is intrinsically truncated, reconstruction may still suffer from noticeable artifacts. To address this, we propose a view-wise projection completion strategy as follows:

1) Extrapolating the truncated projection to half-scan.

The missing data within the half-scan can be efficiently estimated using the mean attenuation coefficients of the scanned object. As illustrated in Fig. 7, the mathematical formulation is given by:

$$P^{v}(i,j) = \overline{u} * L(i,j) \tag{8}$$

where $\overline{u} = \dfrac{\sum_{i=1}^{N_{view}} \sum_{j=1}^{m} P^{r}(i,j)}{\sum_{i=1}^{N_{view}} \sum_{j=1}^{\mathrm{m}} L(i,j)}$, i=1,..., $N_{view}$, $\mathrm{j} = m,...,n$ , $m$ is the quantified detector length and $n$ is the half-scan length of the detector, respectively. $L(i,j)$ is the intersection length of the ray (corresponding to the $j$-th detector unit in the $i$-th view) with the object.

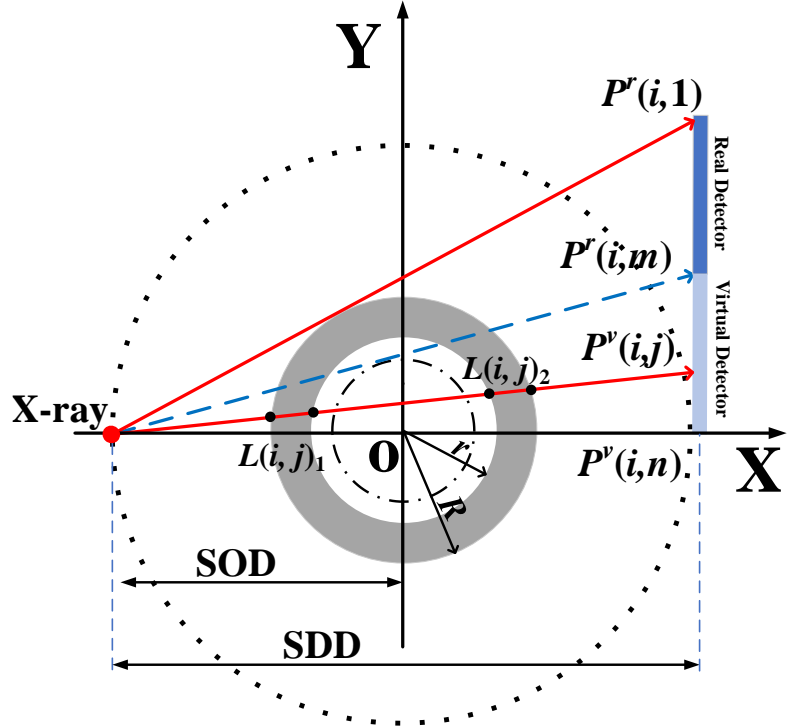

Fig. 7. Completion of tangential scan to half-scan.

2) Mirroring the half-scan to the full-scan. According to projection symmetry in a circular trajectory, the full-scan can be obtained as:

$$\begin{cases} P_1(-\gamma,\beta) = P_2(\gamma,\beta+\pi-2\gamma), & \beta \in (0,\pi) \\ P_3(-\gamma,\beta) = P_1(\gamma,\beta-\pi-2\gamma), & \beta \in (\pi,2\pi) \end{cases} \tag{9}$$

where $\gamma$ is the X-ray angle in the half fan-beam, $\beta$ is the source rotation angle.

The completion process is shown in Fig. 8. Since both $\gamma$ and $\beta$ are discretely sampled in practice, the symmetric source rotation angle $\beta+\pi-2\gamma$ is obtained via linear extrapolation from the adjacent sampled angle $\beta_i$ ($i = 1,\ldots,N_{\mathrm{view}}$). Similarly, the ray angle $\gamma$ is approximated using adjacent sampled values.

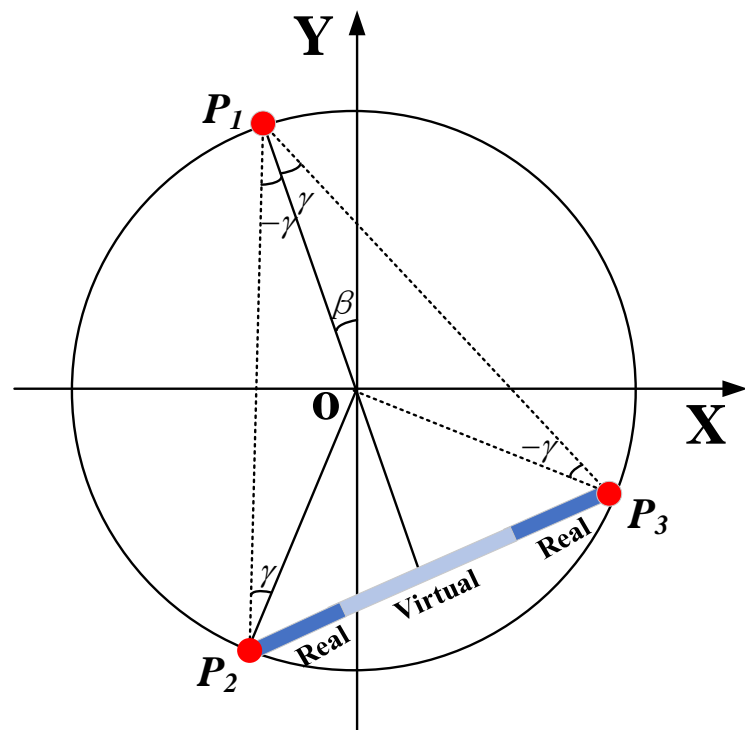

Fig. 8. Completion of half-scan to full-scan.

3) Constraining the extrapolation error. Because the data estimated using the mean attenuation coefficient in Step 1 are relatively coarse, the completed data are further constrained using the projection consistentcy condition [23]:

$$\mathrm{const} = \int_{\pm\gamma_m}^{\pm\gamma_{\max}} P^{r}(\gamma,\beta_i)d\gamma + \int_{-\gamma_{\mathrm{m}}}^{\gamma_{\mathrm{m}}} (P^{v}(\gamma,\beta_i) + \alpha \cdot L(\gamma,\beta_i))d\gamma \tag{10}$$

where $P^{r}(\gamma,\beta_i)$ is the collected data, $P^{v}(\gamma,\beta_i)$ is the estimated data by Eq. (8). $\alpha$ is the constraining parameter, $L(\gamma,\beta_i)$ is the intersection length of ray with object.

Equation (10) enforces consistency among projections acquired from different views. This constraint helps preserve object-specific information in the estimated data, thereby reducing reconstruction errors. Once the complete projection data for TCT is obtained, an initial image can be reconstructed using the ATV model as defined in Eqs. (4)-(5). To further suppress radial artifacts originating from different angular directions, we introduce region-specific gradient weights. Specifically, we apply horizontal and vertical gradient weights to four designated regions ($[\frac{\pi}{4},\frac{3\pi}{4})$, $[\frac{3\pi}{4},\frac{5\pi}{4})$, $[\frac{5\pi}{4},\frac{7\pi}{4})$, $[-\frac{\pi}{4},\frac{\pi}{4})$ ), using values such as (0.6,0.4) or (0.4,0.6), depending on their spatial orientation.

### *2.3. Image artifact suppression network*

Due to residual errors introduced during projection completion (as discussed in Section 2.2), radial artifacts persist in the initially reconstructed images. To address this, we propose an Image Artifact Suppression Network (IAS-Net), illustrated in Fig. 3(b). Because CT images contain abundant nonlocal features [24], the IAS-Net consists of a SWin-transformer-enhanced Encoder-Decoder (SWED) structure, as shown in Fig. 3(e). SWED contains seven Swin Transformer blocks: three in the encoder, three in the decoder, and one at the bottleneck connecting the two parts. The convolutional layers use a kernel size of 3×3. The encoder aims to extract features in the low-dimensional manifold by gradually shrinking the feature maps, identifying artifact features, and being robust to perturbations and noise [25]. The decoder focuses on generating artifact-free results from the encoder outputs by incrementally enlarging the feature maps. In addition, the skip connections between the encoder and decoder combine low-level, fine-grained features with high-level, abstract features, which is beneficial for preserving image details and enhancing information flow [26,27]. In contrast to the traditional encoder-decoder network [28], SWED uses Swin Transformer blocks [29] to capture latent features in long-range dependencies, and employs a shifted local window strategy for efficient computation (as shown in Fig. 3(f)). These strategies are particularly well suited for CT images with nonlocal features. Another difference is that patch merging and patch expanding are used instead of max/mean pooling and transposed convolution. These operations can capture more latent information and avoid the checkerboard effect. To further focus on multidirectional contours, the Asymmetric Convolution Blocks (ACB) [30] are introduced to the encoder as:

$$ACB(I)=\text{Concate}(k_1(I),k_2(I),k_3(I)) \tag{11}$$

where $I$ is the feature map, $k_1$, $k_2$ and $k_3$ are the convolution kernels with sizes of 1×3, 3×3, and 3×1, respectively.

### *2.4. Projection Fidelity Constraint Module*

Measurement fidelity has been shown to be effective in reducing artifact generation and preserving image details in deep learning-based reconstruction methods [31]. To incorporate this principle, we apply a PFC module to the output of IAS-Net, as illustrated in Fig. 3(c). The PFC process proceeds as follows: First, the output image from IAS-Net is forward projected using the same geometric configuration as the original tangential projection acquisition. Then, the residual between this simulated projection and the collected projection data is computed. This residual projection is subsequently reconstructed using a conventional analytical method, such as FBP or FDK, to ensure computational efficiency. Finally, a non-negativity constraint is imposed on the resulting residual image, which helps restore fine contour details and suppress negative artifacts. The process is shown as:

$$I_{\text{res}}=\text{Rec}(P^{r}-\text{Proj}\ (I_{\text{R1}})) \tag{12}$$

$$I_{\text{R2}}=\text{Max}(I_{\text{res}},0)+I_{\text{R1}} \tag{13}$$

where $P^{r}$ is the measured projection, $I_{\text{R1}}$ is the output of IAS-Net.

### *2.5. Image Refinement Network*

Since the reconstruction of the residual projection in Section 2.4 may introduce secondary artifacts in the image, and the multi-step cascaded artifact suppression strategy shows good performance in CT incomplete projection reconstruction [32], we propose an image refinement network (IR-Net) to refine the result of the PFC module, as shown in Fig. 3(d). IR-Net also uses the SWED framework (i.e., the structure of IR-Net is the same as that of IAS-Net), but its purpose is different from that of IAS-Net. Here, IR-Net aims to refine images containing secondary artifacts rather than the radial artifacts targeted by IAS-Net. The training data and network weights are not shared between the two networks.

### *2.6. Loss Function*

An appropriate loss function can accelerate convergence and improve the performance of the deep network [33]. Therefore, a composite loss function is designed for IAS-Net and IR-Net, which consists of $L_2$, $L_1$ norms and structural similarity (SSIM) loss functions to reduce pixel-wise errors, preserve high-frequency edges, and improve structural fidelity, as:

$$L_{\text{comp}} = L_2(I_{\text{roi-R}}, I_{\text{roi-GT}}) + \lambda L_1(I_{\text{roi-R}}, I_{\text{roi-GT}}) + \omega L_{\text{SSIM}}(I_{\text{roi-R}}, I_{\text{roi-GT}}) \tag{14}$$

where $I_{\text{roi-R}}$ is the output of the network and $I_{\text{roi-GT}}$ is the ground truth image. It should be noted that the loss is computed only within the region of interest (ROI), i.e., the annular region. $\lambda$ and $\omega$ are the weighting parameters of the loss terms to ensure that the loss terms are of the same order of magnitude. The $L_2$ loss function is the mean squared error of two images as:

$$L_2(I_{\text{roi-R}}, I_{\text{roi-GT}}) = \frac{1}{N} \| I_{\text{roi-R}} - I_{\text{roi-GT}} \|_2^2 \tag{15}$$

where $N$ is the number of pixels.

The $L_1$ loss function is the mean absolute error of two images as:

$$L_1(I_{\text{roi-R}}, I_{\text{roi-GT}}) = \| I_{\text{roi-R}} - I_{\text{roi-GT}} \|_1 \tag{16}$$

The $L_{\text{SSIM}}$ loss function is the SSIM loss [34] between two images, as presented by:

$$L_{\text{SSIM}}(I_{\text{roi-R}}, I_{\text{roi-GT}}) = 1 - \text{SSIM}(I_{\text{roi=R}}, I_{\text{roi-GT}}) \tag{17}$$

## **3. Experiments and results**

In this study, we evaluated the proposed method on three datasets and compared it with five representative methods: the conventional FBP algorithm [35]; the SART+ATV method [36], an iterative algorithm tailored for incomplete projection data; the Dual-Swin [24], a dual-domain deep network designed for truncated projection reconstruction; the HAL-UNET++ [15], a specialized deep U-shaped network for industrial sparse-view CT reconstruction; and RegFormer [16], an unrolled iterative reconstruction network with learnable regularization. For quantitative evaluation, we adopted three widely used image quality metrics: structural similarity (SSIM) index, the peak signal-to-noise ratio (PSNR), and the root mean squared error (RMSE). Notably, the SART+ATV method only utilized region of interest (ROI) measured projections as input [36], consistent with its original design. For FBP, the input consisted of quantified but uncompleted projections, enabling us to assess the impact of the proposed projection completion strategy. For Dual-Swin, HAL-UNET++ and RegFormer, the inputs were the same as those used by ProTCT, i.e., the quantified and completed projections.

All deep learning models were implemented using the PyTorch framework and optimized using the Adam optimizer. Each dataset was split such that 10% were reserved for testing, while the remaining 90% were randomly partitioned into training (80%) and validation (20%) subsets. Experiments were conducted on a workstation equipped with an Intel Core i7-10700F CPU (2.9 GHz, 16 cores), an NVIDIA GeForce RTX 3060 GPU (12 GB memory), and 32 GB RAM.

### 3.1. Numerical Phantom Results

To verify the proposed method theoretically, we generated 10,000 numerical 3D annuli with inner and outer diameters of 230 and 470 pixels, respectively. Rectangular cracks ranging from 10×10 to 30×30 pixels were randomly intersected with each annulus. Ideal projections and reconstructions were simulated using the ASTRA Toolbox [37] with a cone-beam geometry, the scanning parameters are listed in Table 1. To simulate the complex noise distribution in real cases, the numbers of incident photons in the Poisson noise model were set to 1.8×10^5, 1.9×10^5, 2.0×10^5, and 2.1×10^5 [38].

Table 1

Parameters of scanning geometry.

| Parameter | Value | Parameter | Value |
| --- | --- | --- | --- |
| SOD (mm) | 1550 | Scanning angle range (rad) | [0,2π) |
| SDD (mm) | 1700 | Angular step of projection (deg) | 0.25 |
| Number of detector units (pixel) | 3072×3072 | Size of reconstruction image (pixel) | 512×512 |
| Size of detector unit (mm) | 0.139 | Pixel size (mm) | 0.12 |

According to the ProTCT framework (Fig. 3), the first step is to quantify the necessary projections, as detailed in Section 2.1. This is achieved by solving the optimization problem in Eq. (6) using the CVX toolbox [39], a widely used software package for disciplined convex programming. It is important to note that the projection density is independent of image size, but is strongly influenced by the image class and its sparsity, which refers to the ratio of non-zero pixels within the image [40]. As a result, images belonging to the same class and exhibiting similar sparsity levels share similar projection density requirements.

To express this property quantitatively, we define the projection density as the relative sampling rate, denoted by $R_{\text{view}}$, and formulated as:

$$R_{\text{view}} = \frac{N_{\text{view}}}{N_{\text{suf}}} \tag{18}$$

where $N_{\text{suf}} = \frac{n}{N_{\text{bin}}}$ is the number of reference projection, which means X-ray measurement is more (or equal) to pixel number $n$, and $N_{\text{bin}} = 2N_{\text{side}}$.

As analyzed above, the reconstructed image was resized to 64×64 pixels for efficient computation. Based on the LACT angular range reported in [41] and extensive experiments, the hyperparameters of the sampling model (Eq. (7)) were set as: $(T_\theta, T_{N_{view}}, T_t, a, b, c) = (90^\circ, 32, 0.999, 1.5, 1.5, 2)$ and the same settings were suitable to other two datasets. By verifying each projection $(\theta, N_{\text{view}})$ in $(T_\theta, T_{N_{\text{view}}})$ that satisfied Eq. (6), the relationships between $(\theta, N_{\text{view}}, t)$ were illustrated in Fig. 9.

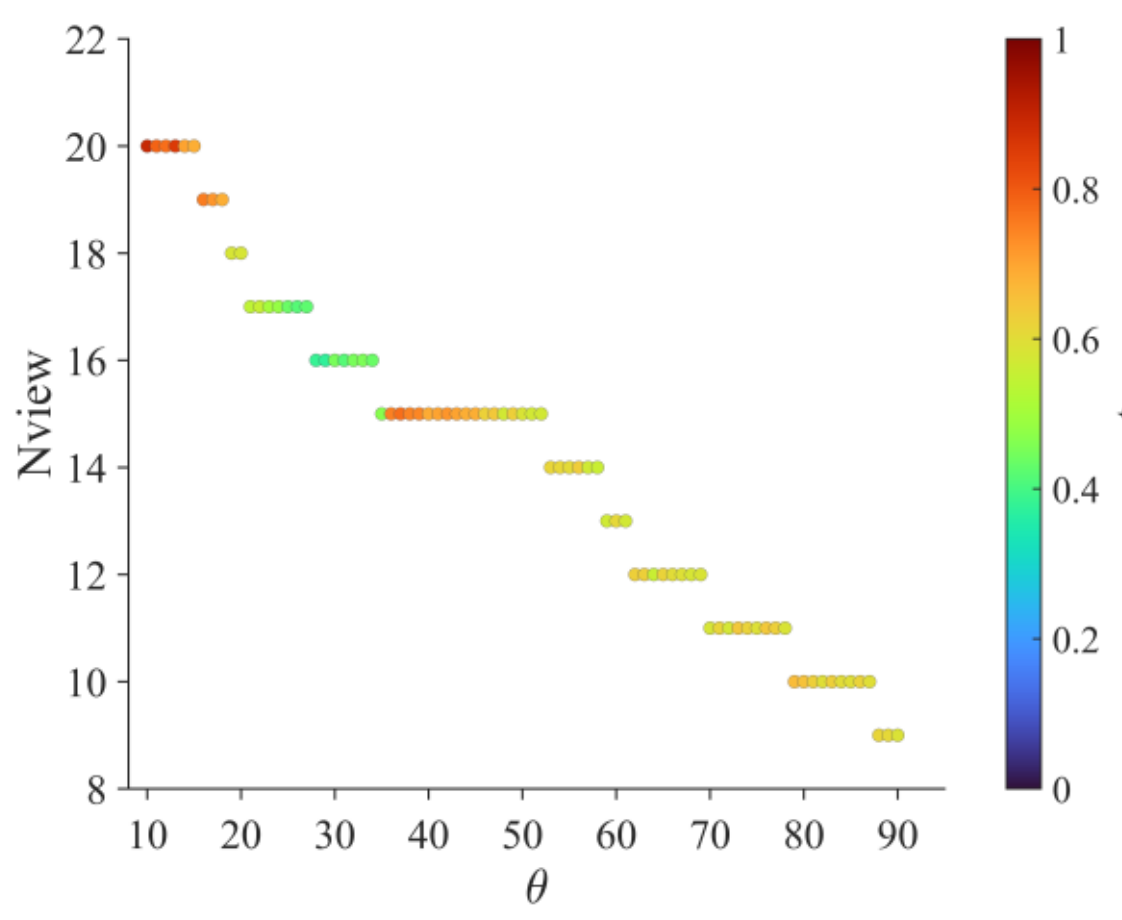

Fig. 9. Relationships among $\theta, N_{\text{view}}$ and $t$ for the numerical phantom.

Fig. 9 shows that as the angular range $\theta$ increases, the number of views $N_{\text{view}}$ and the parameter $t$ gradually decrease. However, in certain angular intervals, $N_{\text{view}}$ remains constant, indicating that image reconstruction also depends on the projection density.

Using Eq. (7), the best sampling conditions for the numerical phantom were $(\theta, N_{\text{view_64}}, t) = (28^{\circ}, 16, 0.379)$. For the original image size $N_{\text{side}} = 512$, the sampling results were $(\theta, N_{\text{view}}) = (28^{\circ}, 128)$. Based on Eqs. (2)-(3), the extended detector length was calculated: $d' = 4$ pixels. After obtaining the necessary projections, the images were initially reconstructed and their artifacts were removed through the process shown in Fig. 3(a)-(d). Specifically, based on extensive experiments, both IAS-Net and IR-Net adopted a learning rate that decayed from 10^-3 to 10^-5, with 100 training epochs and a batch size of 4. The weights of the loss functions were set to $\lambda_{\text{IAS-Net}} = 80$, $\lambda_{\text{IR-Net}} = 20$ and $\omega_{\text{IAS-Net}} = 200$, $\omega_{\text{IR-Net}} = 50$ to ensure that the loss terms have the same order of magnitude in their respective networks.

For a visual comparison of different reconstruction methods, Fig. 10 presents the reconstructed images and corresponding ROIs obtained using FBP, SART+ATV, Dual-Swin, HAL-UNET++, RegFormer, and the proposed ProTCT. It can be observed that the FBP reconstruction suffered from severe artifacts. Focusing on ROI-A, where the number of acquired projection views is significantly lower than in the outer regions, the edges in the FBP result are heavily distorted. In contrast, the artifacts are effectively constrained in the SART+ATV method by regularizing the weighted gradient magnitudes. The DL-based methods substantially improve the reconstruction quality. Specifically, in the Dual-Swin result, the edge structures are clearly recovered; however, some contour erosion is still present and details are lost, as indicated by arrow “1”. This limitation arises because the dominant artifacts in the TCT setting are primarily radially distributed, whereas Dual-Swin does not explicitly enhance directional artifact suppression. In the reconstruction obtained with HAL-UNET++, the visual quality is improved but some edges are over-smoothed due to its reliance solely on gradient-domain sparsity constraints, as indicated by arrow “2”. For RegFormer, the reconstruction quality is further improved by the unrolled iterative scheme that incorporates both a data fidelity term and learnable regularization. Nevertheless, slight edge erosion along the radial direction can still be observed, as indicated by arrow “4”, which may be attributed to the use of square convolution kernels, whereas TCT artifacts typically exhibit radial characteristics. In the region with relatively sufficient projections, such as ROI-B, abrupt artifacts appear in the Dual-Swin result, as indicated by arrow “3”. By contrast, such artifacts are not observed in the other DL-based methods. This may be attributed to the incorporation of data fidelity terms and image-domain constraints into these reconstruction frameworks. Overall, the reconstruction obtained by the proposed ProTCT is visually closest to the reference image, and the edges can be well preserved.

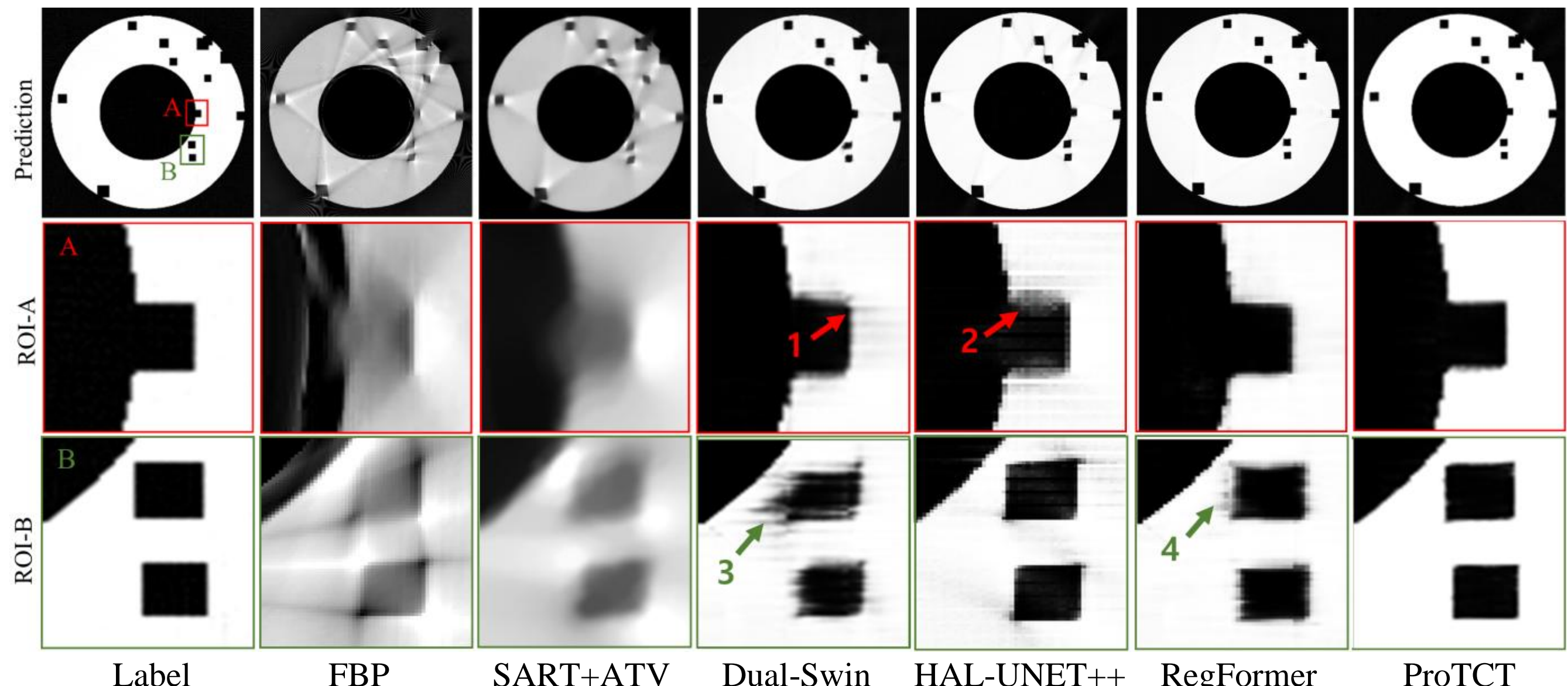

Fig. 10. Reconstruction results of different methods on numerical dataset. The 1st -7th columns represent the images of Label, FBP, SART+ATV, Dual-Swin, HAL-UNET++, RegFormer and ProTCT. The 2nd and 3rd rows show the corresponding ROIs. The display window is [0, 1].

For quantitative evaluation, the means and standard deviations (STDs) of RMSE, PSNR and SSIM are listed in Table 2 (with 2.0×10^5 incident photons). The bold values represent the best results among them. The statistical results indicate that FBP performs worst on the tangential projection, whereas SART+ATV achieves better results. The DL-based reconstruction methods substantially improve all three metrics. Among them, ProTCT achieves the best overall quantitative performance, yielding the lowest RMSE and the highest PSNR and SSIM values.

Table 2

Quantitative evaluations (means±STDs) of different methods for numerical data.

| Method | RMSE | PSNR (dB) | SSIM |
|---|---|---|---|
| FBP [35] | 0.3321±0.0264 | 9.57±0.68 | 0.4756±0.0286 |
| SART+ATV [36] | 0.0501±0.0129 | 22.31±0.48 | 0.6579±0.0203 |
| Dual-Swin [24] | 0.0382±0.0092 | 49.15±0.29 | 0.9538±0.0124 |
| HAL-UNET++ [15] | 0.0312±0.0057 | 49.37±0.26 | 0.9569±0.0173 |
| RegFormer [16] | 0.0251±0.0089 | 50.75±0.23 | 0.9743±0.0157 |
| ProTCT | **0.0242±0.0052** | **53.70±0.17** | **0.9807±0.0084** |

*3.2. Simulated Data Results*

To evaluate the proposed method on data with more complex structures, we adopted the open-source industrial CT (ICT) dataset [42], which contains 4, 000 images of 512×512 pixels. An annular region was extracted from each ICT image to serve as the reconstruction target. The simulations were conducted under the same geometric configuration detailed in Table 1, with Poisson noise incorporated to mimic realistic acquisition conditions, consistent with previous experiments. Similar to the numerical experiment, the relationships between $(\theta, N_{\mathrm{view}}, t)$ are illustrated in Fig. 11.

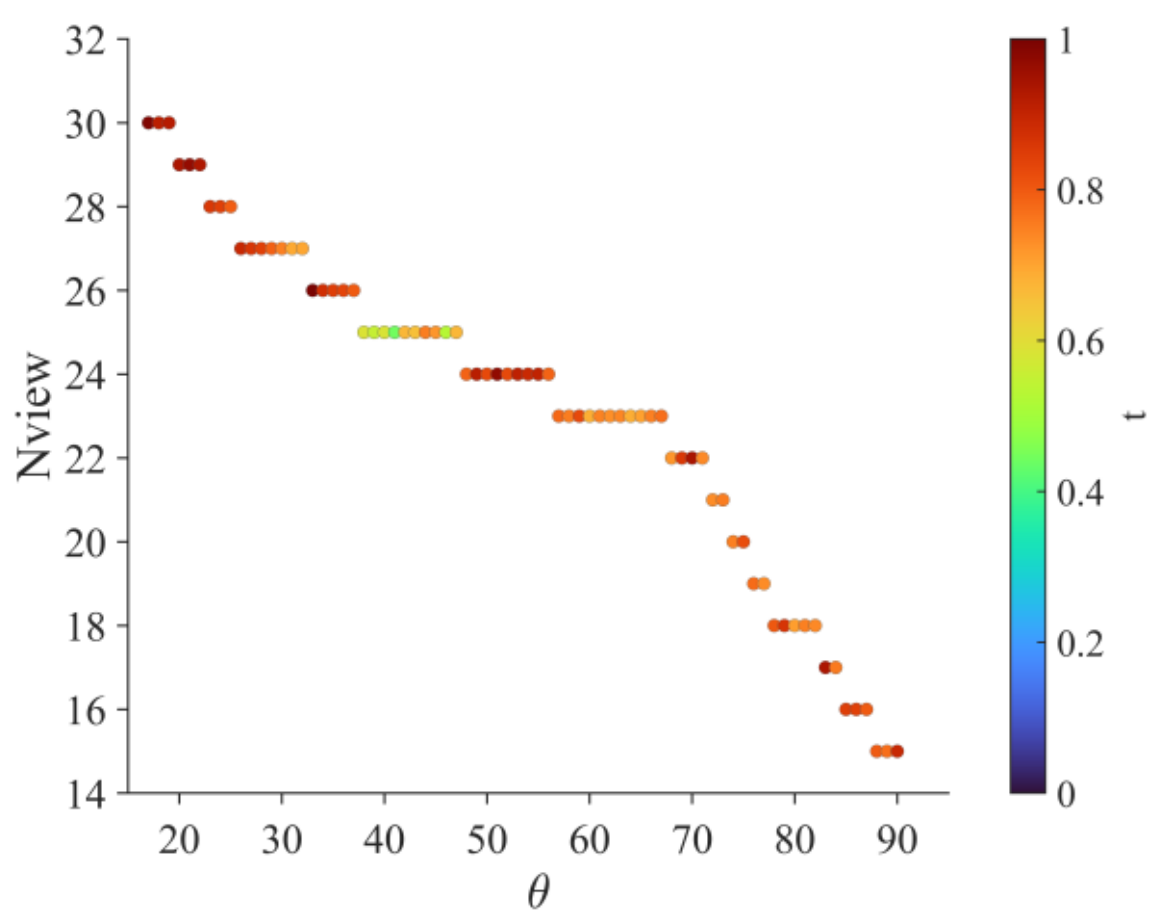

Fig. 11. Relationships among $\theta, N_{\text{view}}$ and $t$ for the simulated data

According to Fig. 11 and Eq. (7), the quantified projections were obtained: $(\theta, N_{view}) = (41^{\circ}, 200)$ and the required detector extension was calculated to be $d' = 8$ pixels. For the DL-based methods, a transfer-learning strategy was adopted, and the models pretrained on the numerical phantom were fine-tuned using the reprocessed ICT image. Specifically, for the ProTCT networks, the learning rates were decreased from 10^-4 to 10^-6, the number of epochs was set to 100, and the batch size was 2. The weights in the loss functions were set to $\lambda_{\text{IAS-Net}} = 0.2$, $\lambda_{\text{IR-Net}} = 0.1$ and $\omega_{\text{IAS-Net}} = 0.5$, $\omega_{\text{IR-Net}} = 0.3$ respectively.

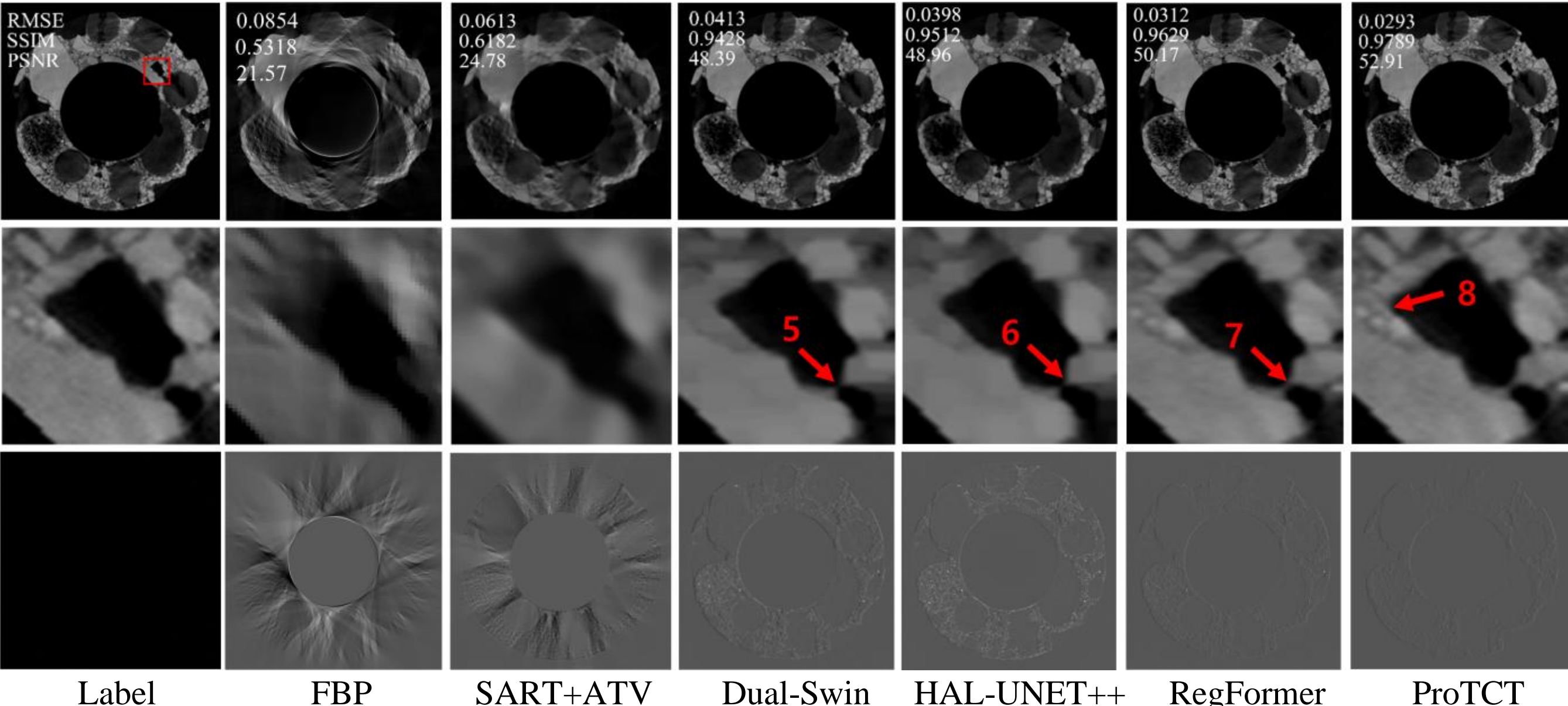

Fig. 12. Reconstruction results of different methods on the simulated dataset. The first row shows the reference image and the reconstruction results obtained using FBP, SART+ATV, Dual-Swin, HAL-UNET++, RegFormer, and ProTCT. The second row shows the corresponding ROIs. The third row shows the difference maps between the reconstructed images and the reference image. The display windows for the images and difference maps are [0, 0.1] and [-0.05, 0.1], respectively.

The reconstructed results obtained using different methods are shown in Fig. 12. It can be observed that the FBP reconstruction suffers from severe artifacts, while SART+ATV produces improved results but still exhibits over-smoothed edges. Among the DL-based methods, most artifacts are effectively suppressed and clearer edge structures are restored. However, some fine details are still missing in the reconstructions of Dual-Swin and HAL-UNET++, as indicated by arrows "5" and "6", respectively. By contrast, RegFormer is able to preserve subtle structural details through its deeply unrolled iterative scheme with explicit measurement consistency, as indicated by arrow "7", achieving performance comparable to that of the proposed ProTCT. Nevertheless, in high-attenuation regions, the proposed ProTCT provides clearer and sharper edges, benefiting from the projection fidelity constraint and the cascaded image refinement network, as indicated by arrow “8”.

In addition, the difference maps between the reconstructions of the comparative methods and the label image are shown in the third row of Fig. 12. These difference maps demonstrate that the proposed ProTCT produces fewer residual errors than the other methods, thereby validating the effectiveness and superiority of the proposed approach.

*3.3. Real projection results*

To evaluate the proposed method under realistic conditions characterized by complex noise distributions [43] and irregular internal structures, an aluminum welded pipe with inner and outer diameters of 94 mm and 106 mm, respectively, was scanned using our in-house-developed CT system. Four representative welding defects were intentionally introduced into the weld at 90-degree intervals, including cracks, porosity, slag inclusion, and lack of fusion [44]. The X-ray source was manufactured by VJ Technologies (USA), model VJ-IXS160BP480P030, with a focal spot size of 0.5 mm. The tube voltage and current were set to 120 kV and 1.2 mA, respectively. The flat-panel detector (FPD) was provided by iRay Technology (China), model iRay-NDT1717M3, with a resolution of 3072 × 3072 pixels and a pixel size of 0.139 mm. A total of 1440 projection views were acquired for high-resolution scanning, and the detailed geometric parameters are listed in Table 1. The acquired projection data were reconstructed using the FDK algorithm implemented in the ASTRA Toolbox, yielding a reconstructed volume of size 1000 × 512 × 512 voxels. Similar to the previous experiments, the relationships between $(\theta, N_{\text{view}}, t)$ are shown in Fig. 13.

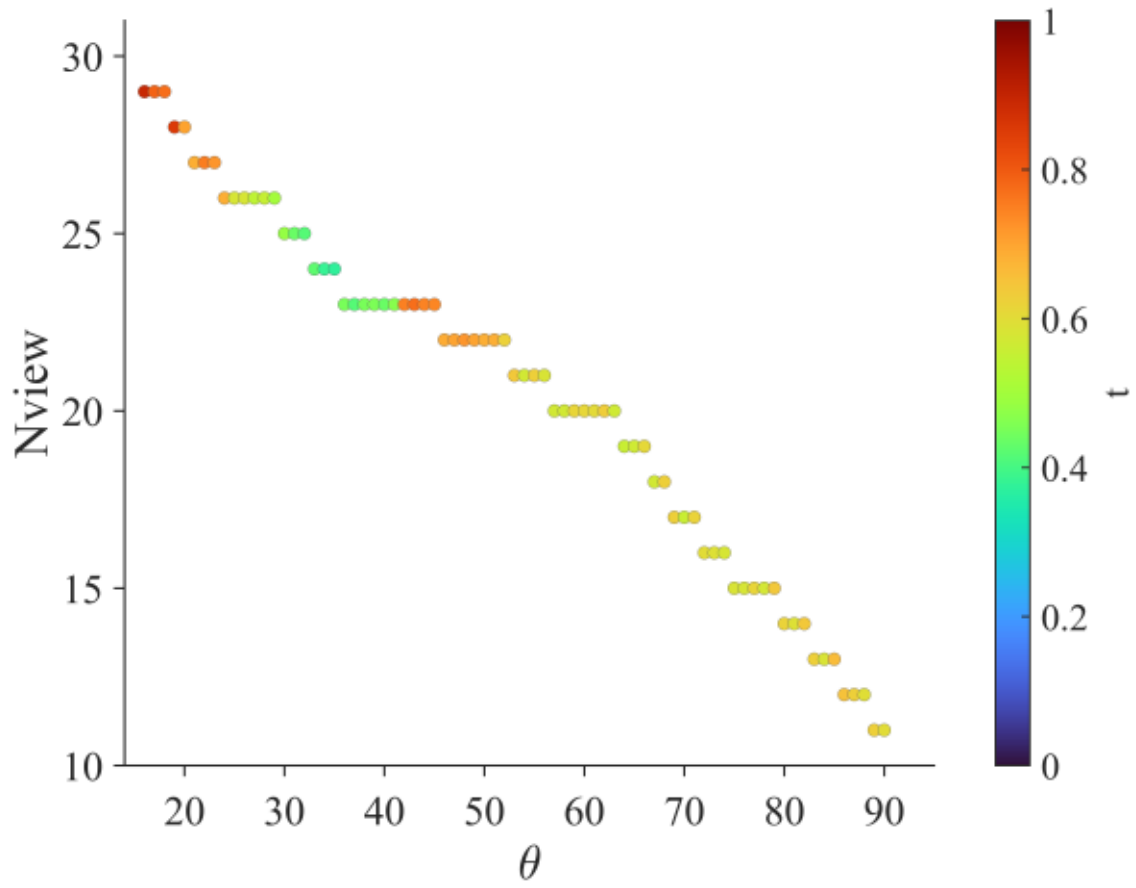

Fig. 13. Relationships among $\theta, N_{\text{view}}$ and $t$ for the real projection

According to Fig. 13 and Eq. (7), the quantified parameters of the welded pipe were determined: $(\theta, N_{view}) = (34^{\text{o}}, 192)$, and the extended length of the detector was $d' = 5$ pixel. The DL models were pretrained on the previous two datasets and were subsequently fine-tuned to accommodate the complex noise and structural variations in the real dataset. Specifically, for the ProTCT networks, the learning rates were decreased from 10^-5 to 10^-7, the number of epochs was set to 100, and the batch size was 2. The weights in the loss functions were set to $\lambda_{\text{IAS-Net}} = 0.0004$ **,** $\lambda_{\text{IR-Net}} = 0.0001$ and $\omega_{\text{IAS-Net}} = 0.01$, $\omega_{\text{IR-Net}} = 0.003$, respectively.

Fig. 14 presents the reconstructed images obtained using different methods. Consistent with the results on the previous two datasets, the FBP algorithm exhibits the worst performance, with structures and details being hardly discernible. The SART+ATV method provides improved restoration but still suffers from obvious over-smoothing. In contrast, the DL-based methods effectively suppress artifacts and yield clearer reconstructions. Specifically, Dual-Swin and HAL-UNET++ are able to recover most of the defect edges in the welded pipe; however, some contours fail to be restored from the degraded projections, as indicated by arrows “9”, “10”, “12”, “13”. Compared with these two approaches, which mainly rely on deep attention mechanisms or gradient sparsity in the image domain, RegFormer employs an unrolled reconstruction scheme incorporating projection consistency and learned local–nonlocal regularization, enabling better recovery of defect structures. Nevertheless, certain subtle details are still lost, as marked by arrow “14”. Overall, among all the compared methods, the proposed ProTCT demonstrates superior performance in suppressing radial artifacts caused by tangential scanning and in restoring fine structural details through the incorporation of measurement consistency and deep attention networks. These advantages can be clearly observed in the enlarged ROIs pointed by arrows “11” and “15”.

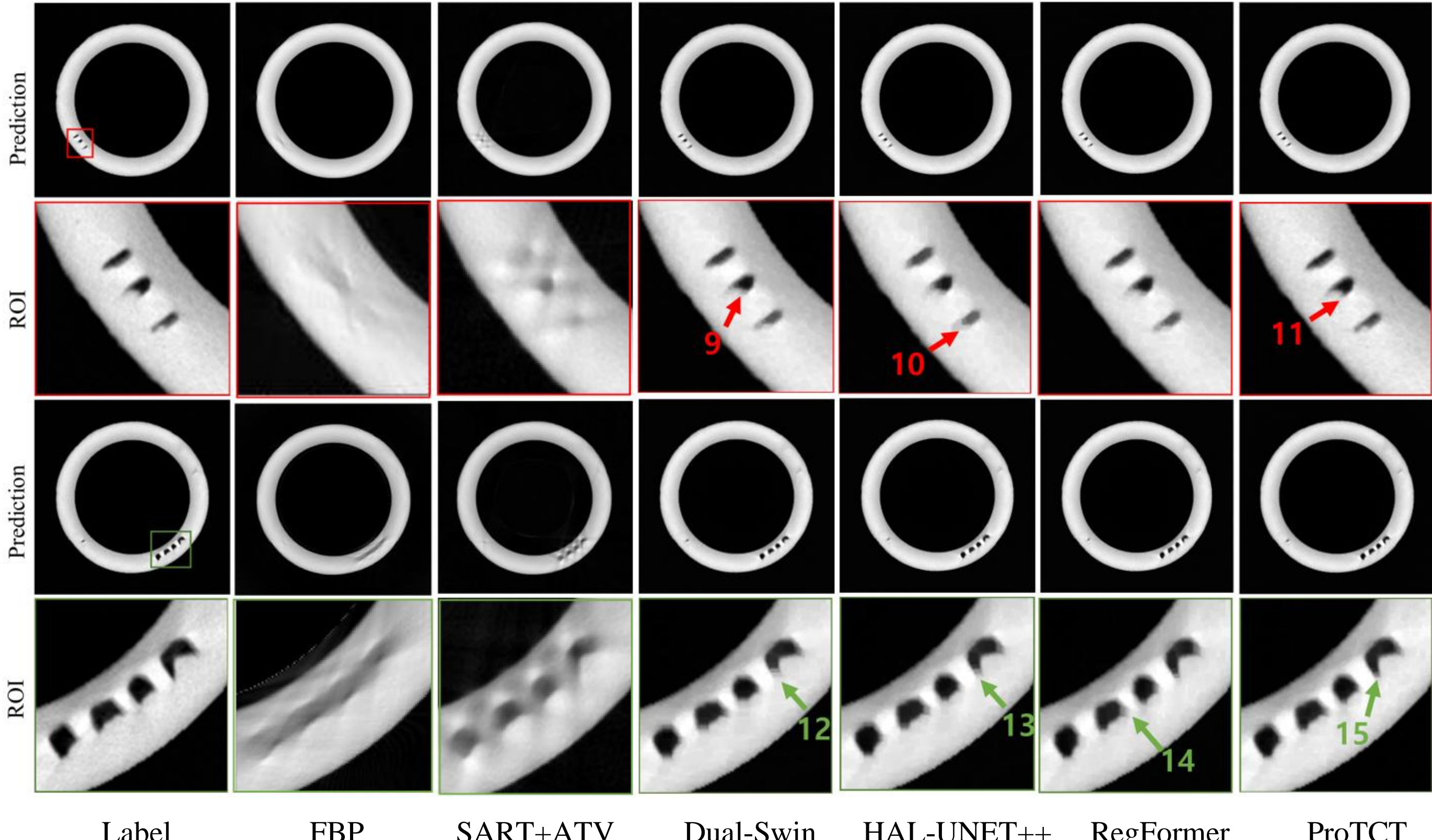

Fig. 14. Reconstruction results of different methods on real dataset. The 1st and 3rd rows represent the images of Label, FBP, SART+ATV, Dual-Swin, HAL-UNET++, RegFormer and ProTCT. The 2nd and 4th rows are the ROIs of the images. The display window is [0.0, 0.03].

Quantitative analysis was also performed for all the methods, and the statistical results are summarized in Table 3. All the DL-based methods achieve substantial improvements over the conventional analytical and iterative approaches. In particular, Dual-Swin and HAL-UNET++ exhibit comparable performance but relatively smaller gains than the other two DL-based methods, indicating the limited capability of projection inpainting and post-processing strategies in addressing tangential CT reconstruction. With the introduction of an explicit measurement-consistency constraint, RegFormer demonstrates more competitive performance than Dual-Swin and HAL-UNET++. Overall, ProTCT obtains the best reconstruction performance in terms of PSNR and SSIM, with RMSE comparable to that of RegFormer. These results demonstrate the effectiveness of the proposed ProTCT in suppressing artifacts and restoring structural details for tangential CT imaging.

Table 3

Quantitative evaluations (means±STD) of different methods for real data.

| Method | RMSE | PSNR (dB) | SSIM |
| --- | --- | --- | --- |
| FBP [35] | 0.1634±0.0078 | 15.48±0.45 | 0.8289±0.0543 |
| SART+ATV [36] | 0.0128±0.0059 | 24.57±0.32 | 0.8763±0.0357 |
| Dual-Swin [24] | 0.0072±0.0008 | 47.32±0.25 | 0.9715±0.0095 |
| HAL-UNET++ [15] | 0.0070±0.0012 | 47.85±0.24 | 0.9796±0.0083 |
| RegFormer [16] | **0.0049±0.0007** | 54.81±0.22 | 0.9813±0.0061 |
| ProTCT | 0.0051±0.0006 | **55.79±0.11** | **0.9972±0.0053** |

Additionally, we analyzed the field of view (FOV) of the proposed method using our in-house-developed CT system as an illustrative example. For conventional full-beam scanning, the FOV diameter is given by:

$$D_1 = 2\mathrm{SOD} \cdot \sin(\arctan(\frac{0.5L_{\mathrm{det}}}{\mathrm{SDD}})) \tag{19}$$

where $L_{\mathrm{det}}$ = 427 mm denotes the detector length, SOD = 1550 mm, SDD = 1700 mm. Under this configuration, the diameter of the conventional full-beam scanning FOV is $D_1$ = 385 mm. When the proposed TCT scanning mode is applied using the same hardware configuration, the FOV diameter $D_2$ can be calculated by referring to Fig. 5 and Eqs. (2)-(3), as:

$$L_{\mathrm{det}} - d' = \frac{D_2 \cdot \mathrm{SDD}}{\sqrt{\mathrm{SOD}^2 - D_2^{\ 2}}} - \frac{r \cdot \mathrm{SDD}}{\sqrt{\mathrm{SOD}^2 - r^2}} \tag{20}$$

where $d'$ denotes the quantified detector extension length calculated using Eqs. (2)-(3), $D_2 = r + d_1$, where $r$ is the inner radius of the annular object and $d_1$ is the wall thickness. The diameter ratio is assumed to be the same as that of the welded pipe used in our experiments. The resulting FOV diameter for TCT scanning is $D_2$ = 1944 mm. Consequently, the achievable imaging diameter is approximately five times that of conventional full-beam scanning.

*3.4. Ablation study*

To investigate the effectiveness of the Projection Quantification and Completion (PQC) module (Fig. 3(a)) and the Projection Fidelity Constraint (PFC) module (Fig. 3(c)) within the ProTCT framework, an ablation study was performed using the numerical dataset. First, the PQC and PFC modules were individually removed from the full ProTCT model to assess their respective contributions. The two modules were then removed simultaneously, leaving only IAS-Net in the ProTCT framework. The quantitative metrics for the numerical dataset are shown in Table 4.

Table 4

Quantitative evaluations for ablation study on numerical dataset.

| Model | RMSE | PSNR(dB) | SSIM |
|---|---|---|---|
| ProTCT | 0.0242 | 53.70 | 0.9807 |
| -PQC | 0.0338 | 31.28 | 0.7352 |
| - PFC | 0.0291 | 46.32 | 0.8527 |
| - PQC - PFC | 0.0362 | 27.57 | 0.6913 |

It can be seen that when either the PQC or PFC module is removed, the quantitative performance degrades. However, the degradation caused by removing the PFC module is less significant than that caused by removing the PQC module. This is because the PFC module mainly preserves image details through measurement consistency, whereas the PQC module completes the projection data and more effectively restores the overall image contours.

In addition, to visually assess the impact of these two modules, the reconstructed results are shown in Fig. 15. As the PQC and PFC modules are progressively removed, the reconstructed images gradually deteriorate. Specifically, when the PQC module is removed, the contours become noticeably distorted, as indicated by arrow “16”. When the PFC module is further removed, the edge details are lost, as indicated by arrow “17”. These observations are consistent with the quantitative results presented in Table 4.

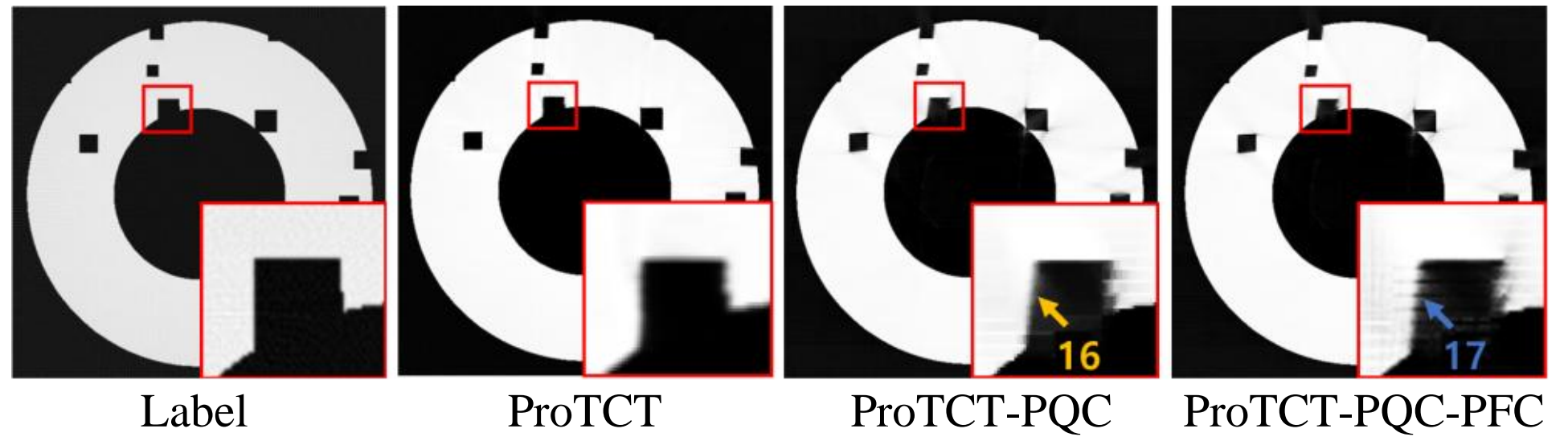

Fig. 15. ProTCT ablation study with different modules. The display window is [0, 1].

### *3.5. Computational cost*

To evaluate the computational cost of different methods, 500 images were selected from the numerical dataset and the metrics are shown in Table 5.

Table 5

Computational cost of different methods (seconds).

| Method | FBP | SART+ATV | Dual-Swin | HAL-UNET++ | RegFormer | ProTCT |
|---|---|---|---|---|---|---|
| Time | 150.7 | 21902.4 | 792.8 | 296.5 | 370.6 | **462.5** |

Table 5 shows that FBP requires the least computation time but produces the lowest-quality images. SART+ATV is the most time-consuming method because of its repeated iterative computations. Dual-Swin takes much less time and achieves better results than the iterative method, HAL-UNET++ is faster than Dual-Swin, as it uses a conventional convolutional U-shaped architecture. In contrast, RegFormer is more time-consuming than HAL-UNET++ because it unrolls the iterative scheme using alternating convolutional and Swin Transformer blocks. ProTCT achieves the best performance while maintaining a reasonable computational cost.

## 4. Conclusion and discussion

This paper proposes a tangential CT reconstruction framework (ProTCT) for imaging large-diameter tubular objects that extend beyond the field of view of a CT system, under the condition that the necessary projections for each object point are physically measured by the detector. The proposed method does not rely on explicit prior assumptions about defect types or shapes. By combining physics-consistent projection information with data-driven compensation, ProTCT is designed to suppress projection-truncation-induced radial artifacts and restore structural details in a general manner. Specifically, we first analyze the scanning scheme of TCT and explore the necessary sampling conditions based on the ATV reconstruction model, which yields an initial reconstructed image. Subsequently, the Swin-Transformer-enhanced encoder-decoder framework is introduced as the backbone to implement a coarse-to-fine reconstruction strategy. To further preserve image details and improve reconstruction reliability, which are critical for deep learning-based reconstruction, a projection fidelity constraint module is incorporated. In the real-data experiment, ProTCT achieved an RMSE of 0.0051, a PSNR of 55.79 dB, and an SSIM of 0.9972. Overall, this work quantifies the sampling conditions of TCT to guide practical detector design and presents a progressive reconstruction framework for improving image quality.

However, although the proposed method shows encouraging performance, several limitations remain and should be addressed in future work. First, in the analysis of the projection for each detector unit, we consider individual “rays” for the theoretical description, whereas in practice, the values registered by each detector pixel include other interactions between radiation and matter, such as scattering, which may introduce additional artifacts. Therefore, introducing anti-scatter grids or estimating the scatter distribution by modeling X-ray photon transport through the object, for example by solving the linear Boltzmann transport equation [45], will be investigated in future work. Second, the proposed quantification method is based on a limited-angle full scan, which is not well suited for TCT because it only satisfies the necessary conditions for TCT reconstruction. Therefore, developing a customized TCT projection quantification model will be a priority in future work. Third, regardless of the accuracy of the quantification method, the missing sinogram data are currently filled using a mean attenuation value for computational efficiency. The exploration of advanced sinogram completion methods may provide better results, such as projection-domain inpainting method [46]. Finally, the reconstruction operator in the PFC module is implemented using an analytical method. Iterative reconstruction methods incorporating geometric prior information could also be employed and may further reduce secondary artifacts.

**Acknowledgments**

The authors would like to thank Prof. Hewei Gao of Tsinghua University for his valuable suggestions and careful review of the manuscript. We also sincerely thank the editors and reviewers for their constructive and insightful comments and suggestions. The authors further acknowledge the Department of Engineering Physics, Tsinghua University, Beijing, China, and the Key Laboratory of Particle & Radiation Imaging (Tsinghua University), Ministry of Education, Beijing, China, for providing support and a favorable working environment during the revision of this manuscript. This work was supported by the National Natural Science Foundation of China under Grant No. 62275223, the School and University Cooperation Project of Sichuan Province (Application No. 25SYSX0134, Project No. 2025YFHZ0032), the Fundamental Research Funds for the Central Universities (Xiamen University, Project No. 20720242006), and XMU Training Program of Innovation and Entrepreneurship for Undergraduates (S202410384330X and 202410384071X).

**Disclosure statement**

The authors report there are no competing interests to declare.

**Data availability statement**

The datasets used or analyzed during the current study are available from the corresponding author on reasonable request.